# Quantum cascade laser frequency stabilisation at the sub-Hz level

Bérengère Argence,[1] Bruno Chanteau,[1] Olivier Lopez,[1] Daniele Nicolodi,[2] Michel Abgrall,[2] Christian Chardonnet,[1] Christophe Daussy,[1] Benoît Darquié,[1] Yann Le Coq,[2] and Anne Amy-Klein[1,*]

[1]*Laboratoire de Physique des Lasers, Université Paris 13, Sorbonne Paris Cité, CNRS, 99 Avenue Jean-Baptiste Clément, 93430 Villetaneuse, France*

[2]*LNE-SYRTE, Observatoire de Paris, CNRS, UPMC, 61 Avenue de l'Observatoire, 75014 Paris, France*

*Corresponding author: amy@univ-paris13.fr*

**Abstract :**

Quantum Cascade Lasers (QCL) are increasingly being used to probe the mid-infrared "molecular fingerprint" region. This prompted efforts towards improving their spectral performance, in order to reach ever-higher resolution and precision. Here, we report the stabilisation of a QCL onto an optical frequency comb. We demonstrate a relative stability and accuracy of $2\times10^{-15}$ and $10^{-14}$, respectively. The comb is stabilised to a remote near-infrared ultra-stable laser referenced to frequency primary standards, whose signal is transferred via an optical fibre link. The stability and frequency traceability of our QCL exceed those demonstrated so far by two orders of magnitude. As a demonstration of its capability, we then use it to perform high-resolution molecular spectroscopy. We measure absorption frequencies with an $8\times10^{-13}$ relative uncertainty. This confirms the potential of this setup for ultra-high precision measurements with molecules, such as our ongoing effort towards testing the parity symmetry by probing chiral species.

Molecules are increasingly being used in precision tests of physics thanks to progress made in controlling molecular degrees of freedom[1,2]. They are now being used, for example, to test fundamental symmetries[3,4], and to measure fundamental constants[5-7] and their possible variation in time[8,9]. Most of these experiments are spectroscopic precision measurements, and are often in the mid-infrared (MIR) domain where the molecules exhibit intense and narrow rovibrational transitions. This creates a need for efficient laser sources in this spectral region, prompting efforts to develop ultra-stable (US) and accurate continuous wave (cw) laser sources as well as MIR frequency combs (see for instance [10,11]). Quantum Cascade Lasers[12,13] (QCL) are promising cw sources, as they are available anywhere in the 3-25 µm MIR range, and each QCL can be tuned over several hundreds of gigahertz. However they have a free-running line width of tens to thousands of kilohertz, making their frequency stabilisation challenging[14-23].

In general, frequency stabilisation of a laser requires one to first choose a frequency reference and then to transfer its spectral properties to the laser. The most common references used in the MIR spectral region are molecular rovibrational absorption lines[3,24]. However, molecular degrees of freedom cannot be controlled as efficiently as atomic ones, leading to limited frequency reproducibility and accuracy. Alternatively, several attempts to develop MIR US cavities have been made, but the performance is far from what is reported in the near-infrared (NIR) or visible spectral regions[25,26].

It is thus appealing to use as frequency references the best US lasers. Those being mostly in the NIR region, one has to bridge the gap between the NIR and MIR domains. This is possible using an optical frequency comb (OFC). The MIR frequency is locked to a high harmonic of the OFC repetition rate, using sum- or difference-frequency generation processes. This not only provides the ultimate stabilities of lasers locked to state-of-the-art US cavities[27,28], but also allows one to benefit from the direct link between such NIR sources and primary frequency standards[29]. A few groups have already demonstrated the stabilisation of a MIR laser to a primary standard traceable frequency reference using an OFC[15-17,19-21,30-32]. Moreover, with the rapid expansion of US frequency transfer over optical fibre links, signals from a NIR US laser referenced to a primary frequency standard can indeed be transferred a few hundreds of kilometers away without any degradation of its stability[33-36]. In the future, the development of a local reference laser may no longer be required.

In this paper we demonstrate the stabilisation of a QCL onto the signal from a remote NIR US laser. This signal is transferred using a 43-km fibre link and its frequency is monitored by primary standards. This leads to a two-order-of-magnitude improvement in frequency stability and traceability over previous work on QCLs[22]. To show the potential for precise spectroscopic measurements, we further demonstrate continuous tuning of this stabilised QCL and measure $OsO_4$ molecular absorption frequencies with a $8 \times 10^{-13}$ uncertainty, well below what has ever been reported with QCLs.

The experimental set-up is displayed in Fig. 1. We use a room-temperature distributed feedback QCL emitting up to 40 mW at 10.3 µm, with a tuning range of 60 nm (see methods). It is phase-locked to an OFC, consisting of an Erbium-doped fibre mode-locked laser emitting around 1.55 µm. The comb repetition rate $f_{\text{rep}}$ ~250 MHz is stabilised by phase-locking one tooth of the OFC onto a NIR frequency reference generated at LNE-SYRTE laboratory and transferred to LPL laboratory through an optical fibre link [33] (see methods). This NIR reference exhibits a relative frequency stability lower than $2 \times 10^{-15}$ between 1 and 100 s. Its absolute frequency is known with an uncertainty of $10^{-14}$ on this time scale when referenced to a H-Maser via a local OFC at LNE-SYRTE (referred to as SYRTE-OFC in Fig. 1 and methods)[29]. The $3 \times 10^{-16}$ Cs fountain's accuracy[29] can be reached for long enough averaging time. At LPL, part of the 1.55 µm comb is amplified and fed into a non-linear fibre to generate an additional output of the OFC, centred at 1.82 µm[31]. After adjusting their shape and polarization, the 1.82-µm comb and QCL beams are overlapped in an $AgGaSe_2$ crystal to perform sum-frequency generation (SFG), resulting in a shifted comb centred at 1.55 µm. This shifted comb is then combined with the 1.55 µm output of the LPL OFC. A fibre delay line is used to overlap the pulses in the time domain. On photodiode PD1 (see Fig. 1), a beat note of frequency $\Delta_1 = \pm [n f_{rep} - \nu_{QCL}]$ is obtained between the QCL frequency $\nu_{QCL}$ and a high-harmonic (of order n) of the repetition rate (see methods). Note that this beat note is independent of the comb carrier-envelope offset frequency $f_0$. This signal at frequency $\Delta_1$ is filtered and mixed with a radio-frequency (RF) oscillator to generate the error signal that drives the phase lock loop used to stabilise the QCL. A bandwidth of several hundreds of kilohertz is obtained. This stabilisation setup can be used on a wide 9-11 µm spectral range and could be extended to any frequency in the whole 3-20 µm range with an appropriate choice of crystal and comb spectrum.

The performance of this set-up is assessed by measuring the relative frequency stability and frequency noise power spectral density (PSD) of the QCL. We first compare the QCL's stability to that of the state-of-the-art secondary frequency standard around 10 µm, a $CO_2$ laser stabilised on an $OsO_4$ saturated absorption line[24,37,38] (hereafter referred to as the "MIR reference", see methods for details). The beat note of frequency $\Delta_2$ between the stabilised QCL and this MIR reference is detected on photodiode $PD_2$ (see Fig. 1) and sent to a frequency counter. The stability of the beat note is the quadratic mean of the MIR reference and the QCL stabilities, and is obtained by calculating the overlapping Allan deviation of the data (see red squares on Fig. 2). The fractional frequency stability equals $5 \times 10^{-14}$ for 1-s averaging time ($\tau$) and decreases as $\tau^{-1/2}$ up to a few tens of seconds. The stability of the beat note of frequency $\Delta_1$' between the MIR reference and the OFC, detected on PD1, is also plotted for comparison (blue circles). The two stabilities are almost identical except for small deviations due to non-stationary effects, and reflect the noise level of the MIR reference. Incidentally, this is the same noise level as was measured previously[31]. The MIR reference contribution being dominant, this measurement can only provide an upper limit on the QCL stability.

The QCL frequency stability is expected to be given by the LPL OFC stability, with only a negligible contribution from the phase-lock loop residual frequency noise. In this section, we evaluate both contributions, one after the other. The OFC stability is assessed by beating an optical mode of the OFC with a second remote US laser located at LNE-SYRTE (US laser #2, see Fig. 1 and methods). The Allan deviation, shown as green triangles in Fig. 2, is lower than $2\times10^{-15}$ from 1 to 100 s and equals the US laser stability[33]. To evaluate the frequency noise contribution from the QCL phase-lock loop, we modify the experimental set-up to simultaneously phase-lock the QCL and the $CO_2$ laser onto the OFC. We then detect the beat note between the OFC-stabilised QCL and $CO_2$ laser. Its stability is a measure of the residual non-common frequency fluctuations between these MIR sources and thus gives an estimation of the frequency noise contribution of the two phase-lock loops. As shown in Fig. 2 (black stars), it is $2\times10^{-16}$ at 1 s averaging time, one order of magnitude below the OFC stability and reaches $10^{-17}$ at 100 s. This demonstrates that the QCL and the $CO_2$ laser frequency fluctuations are identical at a level well below the OFC stability and that both copy the OFC spectral properties. This stability is 10 times better at 1 s than any other MIR laser to date[38] and constitutes an improvement of a factor at least 30 over the previous record with a QCL[22].

To further characterise the QCL stabilisation, the frequency noise PSD of the aforementioned OFC-US laser #2 beat note signal was also measured. The result is displayed in Fig. 3 (green curve c) after being scaled to 10.3 µm (see methods). The frequency noise PSD is $10^3$ Hz$^2$/Hz at 100 kHz Fourier frequency (limited by the approximately 500 kHz locking bandwidth) and exhibits a plateau at $10^{-2}$ Hz$^2$/Hz between 1 and 100 Hz. Fig. 3 also shows the frequency noise PSD of the 43-km noise compensated fibre link[31] (brown line d). It is below or equal to the OFC PSD. As such, the link adds a minor contribution to the OFC noise. Next, Fig. 3 displays the frequency noise PSDs of the beat notes at frequencies $\Delta_2$ between the MIR reference and the OFC-stabilised QCL (red curve a) and $\Delta'_1$ between the MIR reference and the OFC (blue curve b). These two curves overlap almost perfectly below 100 kHz Fourier frequency. They are at a level of 10 Hz$^2$/Hz from 1 Hz to 1 kHz and increase to $10^3$ Hz$^2$/Hz at 100 kHz. The noise level of these PSDs is the sum of the noise contributions from each laser. At low Fourier frequencies (below 1 kHz), the MIR reference noise dominates over the OFC noise, as previously measured in [31]. The bump around 400 Hz comes from the MIR reference locking bandwidth. Above 1 kHz, PSDs (a) and (b) overlap with the OFC noise PSD. Finally, we note a bump around 500 kHz on PSD (a), which corresponds to the QCL locking bandwidth. These measurements confirm that the QCL noise is indeed copying the OFC noise. Fig. 3 also displays the free-running QCL PSD (black curve e) for comparison, highlighting a 12 orders of magnitude reduction in the QCL frequency noise at 1 Hz.

The relevant parameter for spectroscopic applications is the emission line width, which we evaluate from the measured OFC frequency noise PSD following the approach given in [39]. The low frequency contribution needed for this calculation is deduced from the 1-s gate time frequency data used for deriving the OFC stability of Fig. 2. The resulting QCL line shape shown in Fig. 4a exhibits a full width at half maximum (FWHM) of 0.2 Hz (for a 10-mHz resolution bandwidth) which is the narrowest line width reported for a QCL by far. It can be compared to the calculated free-running line width of 300 kHz (found using a resolution bandwidth of 1 kHz). Fig. 4b also shows the beat note signal $\Delta_2$ between the phase-locked QCL and the MIR reference, recorded with a resolution of 125 mHz with a Fast Fourier Transform analyser. A lorentzian fit gives a FWHM of 10 Hz. Although dominated by the MIR reference contribution, such a narrow beat note had never been recorded with a QCL.

The frequency accuracy, or more specifically the frequency traceability to the primary standards of LNE-SYRTE[29], is ensured by the use of phase-lock loops for the QCL frequency stabilisation. The QCL absolute frequency is thus known with an uncertainty of about $10^{-14}$ after 100 s averaging time, when the NIR laser frequency is referenced to the H-maser only. By averaging for long enough, the $3\times10^{-16}$ Cs fountain's accuracy[29] can ultimately be transferred to the NIR reference and in turn to the QCL frequency.

Such an accurate and ultra-stable QCL is ideal for carrying out molecular spectroscopy at the highest resolutions in the MIR spectral region. As a first demonstration, we perform saturated absorption spectroscopy of the $OsO_4$ molecule at an unprecedented resolution and accuracy for QCL sources. $OsO_4$ is a test molecule for high-precision frequency measurements in the MIR region as many tens of its lines have been accurately measured to serve as secondary frequency standards[40,41]. The QCL beam is sent through an $OsO_4$-filled Fabry-Perot (FP) optical cavity. Saturated absorption spectra are recorded in transmission of this cavity by scanning the RF oscillator used to phase lock the QCL onto the OFC (see methods). This allows the laser to keep its extremely high spectral purity, frequency stability and accuracy while being swept.

Figure 5 shows two $OsO_4$ lines in a spectrum spanning 5 MHz in the vicinity of the $CO_2$ R(14) 10 µm emission line. The QCL beam is frequency modulated at 9.5 kHz and third harmonic detection is carried out. The line on the left is the unidentified (i.e. not assigned unequivocally to a molecular transition) $^{190}$OsO$_4$ reference line in the OsO$_4$ absolute frequency grid[40]. Its frequency is reported to be $\nu_{OsO4/R(14)}$ = 29 137 747 033 333 (40) Hz[40]. To our knowledge, this is the first time the other line (on the right) is measured. The inset in Fig. 5 is a 200-kHz span spectrum as typically used to determine line centre absolute frequencies. It shows a SNR of about 200 for a total recording time of 400 s and peak-to-peak linewidth of 25 kHz. Data are fitted to a combination of a third and a fifth derivative of a lorentzian, to account for the line shape deformation caused mainly by the choice of frequency modulation parameters. This simple model allows us to extract the absolute frequency of the line centre with a typical 50-Hz uncertainty given by the nonlinear regression, of the order of the standard deviation of independent measurements (47 Hz for instance for 6 measurements of the reference line).

Exhaustive studies of systematic effects affecting OsO$_4$ line shifts have already been carried out[24]. In this work, care was taken to be in similar experimental conditions as those reported in the literature for the reference line to allow the comparison of absolute frequencies. We evaluate the frequency of a given line as the weighted mean frequency of independent measurements (see methods). The $^{190}$OsO$_4$ reference line frequency is found to be $\nu_{OsO4/R(14)}$ - 9 Hz , with a standard uncertainty of the mean of 22 Hz. The frequency of the unreported line shown in Fig. 5 is measured to be $\nu_{OsO4/R(14)}$ + 4 147 399 (23) Hz. The absolute frequencies of two other lines have been measured and are listed in Table 1. Our results, in very good agreement with those reported in the literature[40-42], lead to fractional uncertainties on absolute frequencies as low as 8x10$^{-13}$. This is one of the lowest uncertainties ever reported using saturated absorption spectroscopy in the MIR region[31,43]. In particular, the uncertainty of the reference line is improved by a factor of 2, demonstrating the potential of our set-up for precision measurements devoted to metrological applications. Owing to the high spectral purity and direct traceability to the LNE-SYRTE frequency standards, this uncertainty is not limited by our ability to control the laser frequency, but by systematic effects such as the pressure shift. Note that thanks to the QCL's tuning range, we were able to record two unreported lines. Our stabilisation setup can straightforwardly be extended to the entire 9-11 µm spectral range without any loss of performance, by using QCLs of adjacent emission spectra.

We have demonstrated the frequency stabilisation of a 10.3-µm QCL on an OFC phase-locked to a remote US NIR frequency reference. We find the stability of the QCL to be at the level of the reference, below 0.06 Hz ($2 \times 10^{-15}$ in relative value) from 1 to 100 s, which is a new record for such lasers. We derive a line width of the order of 0.2 Hz, which to our knowledge makes our QCL the narrowest to date. Furthermore, our set-up allows for frequency traceability to the primary standards, so the QCL absolute frequency is known to within an uncertainty of $10^{-14}$ after 100 s (given by the LNE-SYRTE H-maser standard). The $3 \times 10^{-16}$ accuracy of the Cs fountains can be achieved by sufficiently increasing integration times[29]. Moreover, the ongoing work on dissemination of an optical reference on a continental scale using Internet fibre networks will eventually enable many laboratories to access an US optical reference. The level of stability and accuracy obtained using this method is better than any other reported in the MIR region to date, whether it be using QCLs or otherwise. It also frees us from having to lock the QCL to any particular reference (another laser or a molecular transition), leading to a considerable increase in the accessible spectral window.

QCLs with performances comparable to the most stable of oscillators pave the way for precise spectroscopic measurements on molecules in the MIR spectral region. Our work demonstrates frequency tuning over 100 MHz, allowing high-resolution saturated absorption spectroscopy of molecular transitions, including those in spectral regions which were previously inaccessible. A much broader range of a few gigahertz would be obtained by scanning the OFC repetition rate. This is achievable by generating a tunable sideband of the NIR frequency reference and locking the comb on it. Moreover, this set-up can be extended to the whole MIR spectral region up to approximately 20 µm simply by adapting the comb output wavelength to the frequency to be measured via sum or difference frequency generation in the appropriate crystal (using for instance orientation-pattern GaAs combined with SFG[44]).

At LPL, this QCL-based spectrometer is critical for our ongoing efforts to make the first observation of parity violation (PV) by Ramsey interferometry of a beam of chiral molecules[3]. In this project, vibrational frequencies of two enantiomers are measured and compared. The expected frequency difference is small, and is predicted to be on the order of $10^{-13}$, at most. Furthermore, the project requires a certain flexibility to be able to switch from one molecular species to another in keeping with the on-going production of novel, promising exotic candidate species. Naturally, these molecules have resonances in different spectral regions, many of which only QCLs can attain. Thus, this particular combination of precision, stability and flexibility of tuning in the mid-IR requires a very particular laser setup, of which the first proof-of-principle is successfully demonstrated by results presented in this paper.

**Methods**

**The quantum cascade laser.** We use a commercial device from Alpes Laser. It can be tuned from 967 to 973 cm$^{-1}$ (corresponding to a tuning range of 60 nm or 180 GHz) by varying the temperature (between 283 K and 243 K) and the driving current. Cooling is ensured by a chilled water circuit and a Peltier module driven by a commercial temperature controller. The threshold current is 570 mA. The QCL can be operated at up to 870 mA and exhibits a maximum output power of 40 mW. It is driven by a homemade low-noise current source resulting in a negligible contribution of the current driver to the free-running frequency noise (below 40 Hz/√Hz down to 1 kHz) [22]. The frequency noise of the free-running QCL is shown in Fig. 3 (black curve e). It is about $10^{10}$ Hz$^2$/Hz at 1 Hz and has approximately a 1/$f$ slope up to 250 kHz, followed by a steeper slope for higher Fourier frequencies.

**LPL OFC stabilisation to a NIR reference.** The NIR frequency reference signal is generated at LNE-SYRTE laboratory. This signal is provided by a 1.54 µm cw laser whose frequency is locked onto an US cavity with the Pound-Drever-Hall method[45]. The relative stability of this laser (US laser #1) is approximately 2x10$^{-15}$ at 1 s averaging time[33]. This stability is transferred to the repetition rate of a local OFC (SYRTE-OFC), by phase-locking one tooth of the comb onto the US laser signal. This enables us to measure its absolute frequency by comparison against a primary frequency standard steered H-maser of LNE-SYRTE[29]. Furthermore, the cw US laser frequency is shifted by an acousto-optic modulator (whose frequency is controlled by the SYRTE-OFC measurement system) so as to remove the slow frequency drift of the US cavity. This leads to an US NIR frequency reference traceable to primary standards (H-maser) with a 10$^{-14}$ uncertainty after 100 s. This reference signal, of frequency $v_{ref}$, is then transmitted to LPL through a 43-km long optical fibre link. An active correction of the phase noise accumulated by light propagation in the optical fibre is implemented[33]. The residual phase noise instability added by the compensated link (detected with full bandwidth) lies below $10^{-15}$ at 1 s averaging time ($\tau$) and decreases with a $\tau^{-1}$ slope[31]. It is below the stability of the US laser. The signal of the US frequency reference is thus transferred to LPL without degradation (up to Fourier frequencies of a few tens of hertz[33]). At LPL, it is used to phase-lock a local 1.54 µm laser diode for signal regeneration. The repetition rate $f_{rep}$ of the LPL OFC is in its turn phase-locked to this laser diode after removal of the comb carrier-envelope offset frequency $f_0$[46]. Note that for this lock we use some small fraction of light at 1.55 µm present in the comb output at 1.82 µm (see next section) in order to take into account the possible phase fluctuations arising from the comb amplifier used to generate this 1.82 µm output[47]. Fast and slow corrections are respectively applied to an intra-cavity electro-optic modulator and a piezo-electric transducer (PZT) acting on the cavity length, resulting in a bandwidth of more than 500 kHz. The comb is then operating in the "narrow-line width regime"[48].

**QCL stabilisation onto the OFC.** The 1.55 µm LPL-OFC is used to transfer the spectral properties of the NIR reference to the QCL. An additional output of the comb centred at 1.82 µm (total power 25 mW, mode frequencies $q\,f_{rep} + f_0$ where q is an integer, $f_{rep}$ is the OFC repetition rate, and $f_0$ is the carrier-envelope offset frequency) is overlapped with the QCL beam (5 mW, frequency $\nu_{QCL}$) in an $AgGaSe_2$ crystal to perform the SFG. A set of lenses is used to focus the QCL beam in the crystal. An iris diaphragm and an optical isolator, composed of a wire-grid polarizer and a quarter-wave plate, minimize the optical feedback to the QCL mainly caused by back-reflection from the $AgGaSe_2$ crystal facets[21]. SFG between the 1.82 µm comb and the 10.3 µm cw beam results in a shifted comb (30 nW, mode frequencies $q\,f_{rep} + f_0 + \nu_{QCL}$), centred at 1.55 µm. This shifted comb is then combined with the 1.55 µm output of the LPL OFC (10 mW, mode frequencies $p\, f_{rep} + f_0$ where p is an integer,) resulting in a beat note of frequency $\Delta_1 = \pm [(p-q)\, f_{rep} - \nu_{QCL}]$ (with p-q = n an integer of the order of 120000), independent of the OFC carrier-envelope offset frequency $f_0$. This beat note exhibits a signal to noise ratio (SNR) of about 25-30 dB in a 100 kHz bandwidth. This SNR is found to be independent of the QCL power since the same SNR is obtained with a higher power (50 mW) $CO_2$ laser[31]. It is limited by the broad frequency noise of the 1.82 µm output originating from the amplified spontaneous emission of the amplifier used to generate this 1.82 µm output. The beat note at frequency $\Delta_1$ is then filtered and mixed with a RF oscillator to generate the error signal that drives a proportional-integrator phase-lock loop. The correction signal is applied to the QCL's current.

**MIR reference.** The MIR reference is a $CO_2$ laser stabilised onto an $OsO_4$ saturated absorption line. This system constitutes the state-of-the-art MIR secondary frequency standard[24,37] in the 10 µm region. In this work, the $CO_2$ laser is stabilised on the $^{190}OsO_4$ reference line in the vicinity of the 10 µm R(14) $CO_2$ emission line[40]. The $OsO_4$ gas fills a 1.5-m long Fabry-Perot cavity which has a finesse of about 100 [37]. For practical reasons, we use a second $CO_2$ laser phase-locked to the $OsO_4$-stabilised $CO_2$ laser. As previously demonstrated[31] and expected from a properly working phase lock loop, no phase noise is added to the second $CO_2$ laser. Note that the experimental set-up allows the $CO_2$ laser to be sent to the SFG crystal (instead of the QCL, see Fig. 1) in order to evaluate the MIR reference frequency noise and stability.

**OFC frequency noise and stability.** In order to obtain the LPL OFC spectral characteristics, we evaluate the stability and the frequency noise of the beat note between an optical mode of this comb and another US laser (after removing the comb offset frequency $f_0$ [46]). This US laser

#2 consists of a 1.54-µm cw source frequency locked onto an US cavity similar to the one used to provide the NIR reference. The two US lasers thus present the same spectral properties. Since they are transferred from LNE-SYRTE to LPL through the same optical fibre link and since their frequencies are very close, only 375 MHz apart, the phase noise accumulated along the fibre link is approximately the same for both lasers. Thus any propagation noise is cancelled in the beat note between the OFC and the second US signal and compensating the link noise is not needed for this measurement. The beat note PSD (green curve c in Fig. 3) resulting from the sum of the OFC and the US laser #2 noises thus gives an upper limit on the OFC frequency noise. The residual noise PSD of the stabilised fibre link (brown curve d in Fig. 3) should be added to it to obtain the overall frequency noise of the OFC. Nonetheless, the link noise contribution is found to be negligible except for Fourier frequencies between 20 and 400 Hz, where it becomes comparable to that of the OFC, leading however to a minor overall contribution. Note that these PSDs have been scaled to 10.3 µm by the square of the wavelengths ratio for a direct comparison with PSDs recorded at 10.3 µm. From the beat note frequency measurement, we also derive the OFC stability by calculating the overlapping Allan deviation of the data and dividing it by $\sqrt{2}$, assuming that the two systems are independent and identical. It is displayed as green triangles in Fig. 2, after removal of the residual linear drift of the cavities (0.3 Hz/s) from the raw frequency data. The compensated link stability is $9 \times 10^{-16}$ at 1s and decreases as $\tau^{-1}$, thus hardly contributing to the OFC stability[31].

**OsO$_4$ spectroscopy.** Saturated absorption spectroscopy is performed inside a FP cavity to increase the SNR[24,37]. The molecular signal is detected in transmission of the cavity. For detection purpose, we apply two frequency modulations to the QCL. The first modulation at 18 kHz with a deviation of about 90 kHz allows the FP cavity to be locked onto the laser using first-harmonic detection. The correction is applied to a PZT on which is mounted one of the FP cavity mirror. The second modulation at 9.5 kHz with a deviation of 36 kHz is used to detect the third harmonic of the saturated absorption signal with a lock-in amplifier. The choice of modulation parameters results from a trade-off between an increase in the SNR and a broadening of the molecular absorption line. Typical experimental parameters are a pressure of 0.04 Pa (corresponding to an absorption of 50 % detected in transmission of the cavity), a power of 50 µW inside the cavity and a beam diameter of 6 mm. The QCL frequency is swept by scanning the frequency of the RF oscillator used to phase lock the QCL onto the OFC. Each spectrum is recorded by scanning the frequency up and down in turns in order to avoid any line shift induced by the sweeping[49]. The spectrum in the inset of Fig. 5 is the result of 10 up and down scans of 200 steps of 1 kHz with a recording time constant of 5 ms per point and 100 ms between points. The total recording time is thus 400 s. Each line frequency is obtained as the weighted mean of the line centres determined from each measurement. The uncertainty on one measurement, given by the nonlinear regression, is a conservative value since the fitting profile does not perfectly match the experimental line shape (as shown by the small remaining residuals). The uncertainty of the mean is evaluated as the experimental standard uncertainty (calculated from the uncertainties of each single measurement). It gives a conservative value of the error since it is typically higher than the standard deviation of the mean. The present set-up allows scanning continuously over about 100 MHz with a limit given by $f_{rep}/2$. To go beyond, it is possible to directly sweep the OFC repetition rate while keeping its spectral purity. Currently, the repetition rate is in fact locked to a local laser diode, itself phase-locked to the NIR reference (see above). By using a wideband electro-optic modulator (EOM), one can generate a sideband to this laser diode and phase-lock it onto the NIR reference. The laser diode frequency, and in turn the OFC repetition rate and the QCL frequency can be swept by scanning the sideband frequency. This way scanning the QCL over 1 GHz is achievable.

**Acknowledgements**

We are very grateful to Paul-Eric Pottie, Giorgio Santarelli, Rodolphe Le Targat and Won-Kyu Lee for fruitful discussions as well as helping us with the US NIR lasers or the active compensated optical link. We also thank Sean Tokunaga for carefully reading the manuscript and for helping us with line width calculations. The authors acknowledge financial support from CNRS, Agence Nationale de la Recherche (ANR BLANC LIOM 2011 BS04 009 01, ANR QUIGARDE 2012 ANR-12-ASTR-0028-03 and ANR NCPCHEM 2010 BLAN 724 3), Labex First-TF (ANR 10 LABX 48 01), AS GRAM and Université Paris 13.

**Author contributions**

BA, BC, OL, YLC, AAK conceived and designed the experiment. BA, BC, DN, BD, AAK performed the experiment. BA, OL, DN, MA, CD, BD, YLC, AAK analysed the data. BA, MA, BD, YLC, AAK contributed materials/analysis tools. All authors discussed the results. BA, OL, DN, MA, CC, CD, BD, YLC, AAK wrote the paper.

**Competing Financial Interests statement**

The authors declare no competing financial interests.

**Additional information**

Correspondence and requests for materials should be addressed to B.A. or A.A.K.

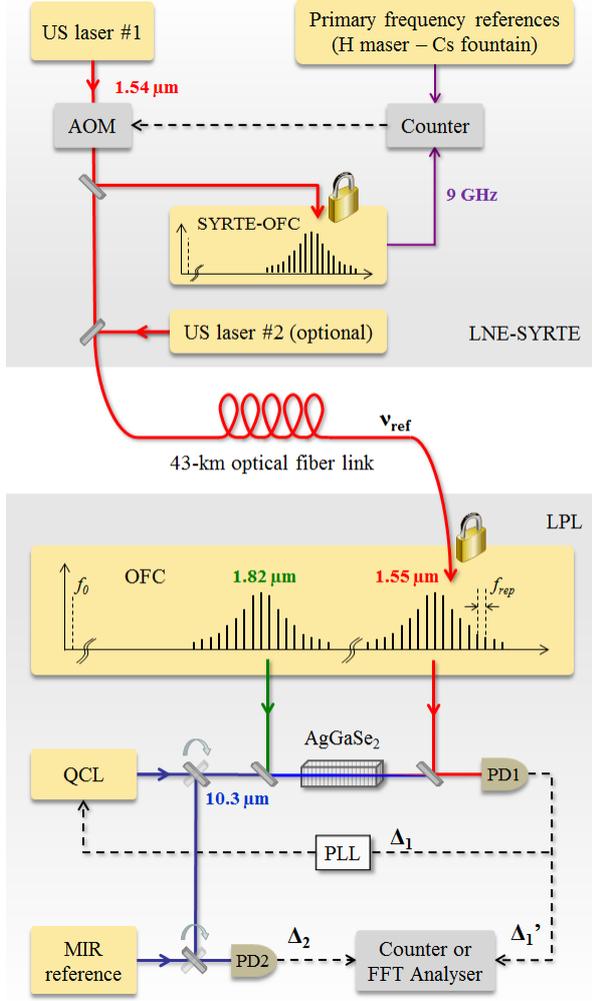

**Figure 1: Experimental setup.** The NIR frequency reference $\nu_{ref}$ generated at LNE-SYRTE with the US laser #1 is transferred to LPL through a 43-km long optical fiber link. Its absolute frequency is measured against primary frequency standards by use of an Optical Frequency Comb (SYRTE-OFC). Its stability is provided by an ultra-stable (US) cavity (see methods) whose frequency drift is removed by use of an acousto-optic modulator (AOM). At LPL, the comb repetition rate $f_{rep}$ of an OFC is phase-locked onto this reference. The Quantum Cascade Laser (QCL) is then phase-locked onto a harmonic of $f_{rep}$ by performing sum frequency generation in an AgGaSe$_2$ crystal (see methods). The beat note of frequency $\Delta_1$ is processed to generate the error signal for the QCL phase lock loop (PLL). The signal of frequency $\Delta_1'$ (resp. $\Delta_2$) on photodiode PD1 (resp. PD2) corresponds to the beat note between the MIR

reference and a multiple of $f_{\text{rep}}$ (resp. the QCL). The stabilities and frequency noise PSDs are evaluated by using a counter and a Fast Fourier Transform (FFT) analyser. The arrows indicate movable optics and the padlocks symbolize the OFCs' PLLs. The US laser #2 is used to evaluate the spectral purity of the LPL comb.

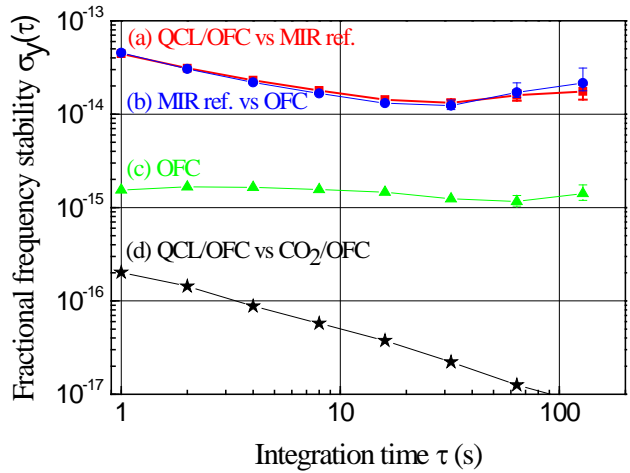

**Figure 2: Fractional frequency stability** of (a) the beat note between the QCL phase-locked onto the NIR reference (QCL/OFC) and the MIR reference (■), (b) the beat note between the MIR reference and the OFC (●), (c) the OFC (▲) and (d) the beat note between the QCL and the $CO_2$ laser, both stabilised onto the OFC (⋆). Overlapping Allan deviations are processed from data measured using a Π-type counter with 1-s gate time. For curve d, a dead-time free counter was used.

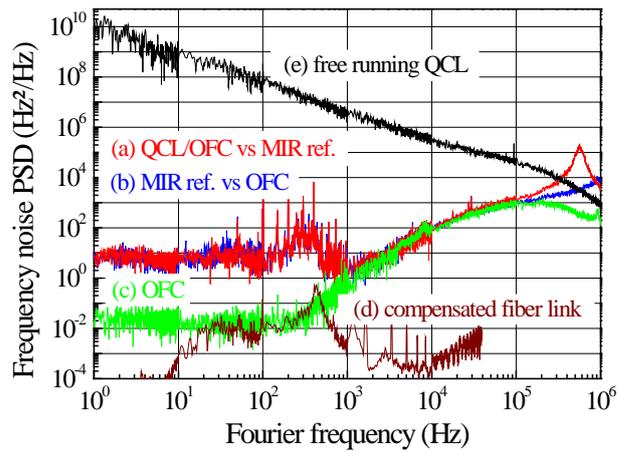

**Figure 3. Frequency noise Power Spectral Density (PSD)** of (a) the beat note between the QCL phase-locked onto the NIR reference (QCL/OFC) and the MIR reference (red), (b) the beat note between the MIR reference and the OFC (blue), (c) the OFC (green), (d) the noise compensated 43-km optical fibre link (brown) and (e) the free running QCL (black). All these PSDs are relative to a carrier frequency of 29.1 THz (10.3-µm wavelength).

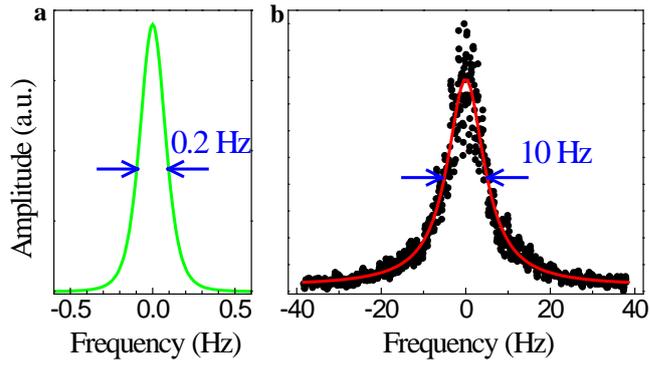

**Figure 4. QCL line shape and beat note with the MIR reference.** a) Calculated QCL line shape showing a 0.2-Hz FWHM (10-mHz resolution bandwidth). This estimation is based on the temporal data used to derive the OFC relative stability (curve c of Fig. 2) and the OFC frequency noise PSD (curve c of Fig. 3). b) Beat note between the QCL phase-locked onto the NIR reference and the MIR reference, recorded with a FFT analyser (125 mHz resolution, average of 10 sweeps of 8 s). The beat note is offset to zero by subtracting approximately 50 kHz. The red line is a lorentzian fit of the data with a line width (FWHM) of 10 Hz.

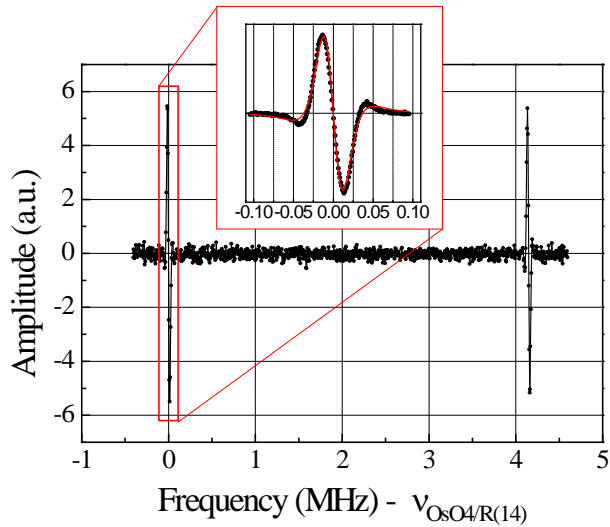

**Figure 5. OsO$_4$ spectrum, in the vicinity of the R(14) emission line of CO$_2$.** Two lines are recorded over a span of 5 MHz by steps of 5 kHz, using third harmonic detection. The x-axis is offset by ν$_{OsO4/R(14)}$ = 29 137 747 033 333 Hz, corresponding to the reported absolute frequency of the OsO$_4$ reference line (left-hand side line)[40]. The right-hand side line has not been reported yet in the literature. The inset shows a spectrum of the reference line recorded over 200 kHz with steps of 1 kHz. These data are fitted to the sum of a third and fifth derivative of a lorentzian.

**Table**

| OsO$_4$ lines in the vicinity of the CO$_2$ R(14) laser line at 10.3 μm | Frequency shift from ν$_{OsO4/R(14)}$ calculated from references[40,42] (in kHz) | Frequency shift from ν$_{OsO4/R(14)}$ measured in this work (in kHz) |
|---|---|---|
| $^{190}$OsO$_4$ reference line (unassigned) | 0.000 (40) | -0.009 (22) |
| Unreported line | - | +4 147.399 (23) |
| $^{190}$OsO$_4$, R(46)A$_1^3$(-) | +101 726.83 (5) | +101 726.821 (32) |
| Unreported line | - | +123 467.401 (32) |

**Table 1. Absolute frequencies of four OsO$_4$ absorption lines in the vicinity of the R(14) CO$_2$ laser line.** The frequencies are given with respect to the OsO$_4$/CO$_2$-R(14) reference line frequency, ν$_{OsO4/R(14)}$ = 29.137 747 033 333 THz, reported in [40]. In the second column we report the absolute frequencies calculated from references[40,42] with 1σ-uncertainty. The third column displays the results of this work, where the uncertainty is the experimental standard uncertainty of the mean. The R(46)A$_1^3$(-) line has previously been recorded at lower pressure[41]. Our measurement is thus expected to be pressure-shifted by approximately +10 Hz[24].